\documentclass{aa}

\usepackage{graphicx}
\usepackage{txfonts}
\usepackage{float}

\usepackage{natbib}
\bibpunct{(}{)}{;}{a}{}{,}

\begin{document}

   \title{Size, atmospheric, and ring constraints on the large Kuiper belt object (55565) Aya from a stellar occultation}

   \subtitle{}

   \author{K. Arimatsu\inst{1,2}\fnmsep\thanks{Corresponding author: ko.arimatsu@nao.ac.jp}
        \and H. Hanayama\inst{3,1,2}
        \and S. Takita\inst{4}
        \and F. Yoshida\inst{5,6}
        \and  T. Hayamizu\inst{7}
        \and J. Watanabe\inst{8,1}
        }

   \institute{
     National Astronomical Observatory of Japan, 2-21-1 Osawa, Mitaka, Tokyo 181-8588, Japan
    \and Ishigakijima Astronomical Observatory, Public Relations Center, National Astronomical Observatory of Japan, 1024-1 Arakawa, Ishigaki,
Okinawa 907-0024, Japan
   \and Department of Physics, Kurume University, School of Medicine, 67 Asahimachi, Kurume city, 830-0011 Fukuoka, Japan
   \and Institute of Astronomy, Graduate School of Science, The University of Tokyo, 2-21-1 Osawa, Mitaka, Tokyo 181-0015, Japan
   \and University of Occupational and Environmental Health, Japan, 1-1 Iseigaoka, Yahata, Kitakyushu, Fukuoka 807-8555, Japan
   \and Planetary Exploration Research Center, Chiba Institute of Technology, 2-17-1 Tsudanuma, Narashino, Chiba 275-0016, Japan
   \and Saga Hoshizora Astronomy Center, 328 Nishiyoka-cho, Oaza Takataro, Saga, Saga 840-0036, Japan
   \and Kyoto Sangyo University, Koyama Space Science Institute,  Motoyama, Kamigamo, Kita-ku, Kyoto 603-8555, Japan}

   \date{Received ...}

  \abstract
   {}
   {We aim to constrain the projected size, geometric albedo, possible atmosphere, and possible ring material of the mid-sized trans-Neptunian object (55565) Aya from the stellar occultation of 2024 May 10 (UT).}
   {We observed the event from three Japanese sites with high-cadence optical imaging. We combined one positive occultation chord with an informative prior on the shadow centre derived from orbital and astrometric uncertainties, and inferred the body radius with Bayesian Markov chain Monte Carlo modelling. We constrained a possible atmosphere by fitting refraction models for pure CH$_4$, N$_2$, and CO atmospheres to multi-band light curves, and searched for ring-like extinction signatures in unbinned and binned high-time-resolution photometry.}
   {We derive an equivalent radius, i.e., the radius of the circle fitted to the projected shadow, of $r=401^{+28}_{-40}$ km for Aya, corresponding to a geometric albedo of $p_V=0.101^{+0.023}_{-0.013}$. This estimate is insensitive to the adopted ephemeris solution and is consistent with an ephemeris-independent analysis that makes no assumption about the position of the shadow centre. We obtain 3$\sigma$ upper limits on the atmospheric surface pressure of $p_0=60$, $90$, and $80$~nbar for pure CH$_4$, N$_2$, and CO compositions, respectively. No significant ring-like extinction is detected over projected distances of about 430--8060 km sampled by the three stations, and our most sensitive data exclude dense narrow features with apparent opacity $p'\gtrsim0.1$ and radial widths of a few kilometres or larger over $\sim3r$--$14r$.}
  {This is the first stellar occultation by Aya ever observed. In addition to constraining Aya's size and albedo and finding no ring-like extinction, we place a stringent atmospheric upper limit that supports diversity in atmospheric states among similarly sized trans-Neptunian objects.}

\keywords{Kuiper belt: individual: (55565) Aya --
                Occultations --
                Techniques: photometric
               }

   \maketitle
   \nolinenumbers

\section{Introduction}
Trans-Neptunian objects (TNOs) in the Kuiper belt present significant observational challenges due to their small angular sizes and great distances.
Stellar occultations provide a complementary and often uniquely powerful technique to overcome these limitations \citep{Sicardy2024-fu}.
When a TNO passes in front of a background star, its shadow sweeps across the Earth.
If the event is recorded with sufficiently high-cadence photometry, the measured disappearance and reappearance times of the star probe the object's instantaneous local limb at a kilometre-level spatial resolution.
Occultation observations can thus provide accurate equivalent radii and projected shapes which, combined with absolute photometry, yield geometric albedos and hence constraints on surface properties \citep{Sicardy2011-dl,Ortiz2012-sc}.
Occultations can also reveal narrow rings or other confined material that is undetectable in direct imaging \citep{Braga-Ribas2014-pz,Ortiz2017-mw,Morgado2023-qj}, and are exquisitely sensitive to tenuous atmospheres, allowing upper limits at the nanobar level \citep{Sicardy2011-dl,Arimatsu2019-sk}. Very recently, evidence for a $\sim$100\,nbar-class atmosphere around the Kuiper belt object (612533)~2002~XV\textsubscript{93} with radius $r\sim300$\,km has been obtained from stellar occultation observations \citep{Arimatsu2026NatAstro}.
Stellar occultation observations thus play a central role in characterising the fundamental properties of TNOs and Centaurs.

(55565) Aya (provisional designation of 2002 AW\textsubscript{197}) is one of the larger mid-sized ($r=200$--$500$\,km) TNOs,
with an orbit in the non-resonant ``hot classical'' Kuiper belt (semi-major axis $\sim 47.3$ au, eccentricity $\sim 0.13$).
The absolute magnitude of Aya is $H \sim 3.6$ mag \citep{Alvarez-Candal2016-sf}, and thermal infrared observations with the Spitzer Space Telescope and Herschel Space Observatory indicate a mean radius of $r \sim 350\mbox{--}400$ km and a relatively high geometric albedo ($p_V \sim 0.1$--$0.2$) compared to typical classical objects, suggesting a bright surface and possibly a complex thermal history \citep{Vilenius2014-wt,Cruikshank2005-ii}.
Rotational light-curve studies have revealed a low-amplitude modulation and suggested a rotation period of approximately $8.8$--$8.9$~h, consistent with a body with small elongation in projection \citep{Ortiz2006-dl,Thirouin2010-lj}.

Since its discovery, Aya has followed an orbit that keeps it significantly away from the Galactic plane for an extended interval.
The object thus has been repeatedly detected serendipitously in a variety of deep optical and near-infrared imaging surveys, including the Near-Earth Asteroid Tracking survey (NEAT), SDSS, and Pan-STARRS1, as well as in targeted follow-up programmes with large telescopes, yielding a long observational arc and a dense set of high-precision astrometric measurements \citep{Pravdo1999-neat,York2000-sdss,Flewelling2020-ps1}.
This extensive astrometric record translates into comparatively small uncertainties in Aya's orbital elements and ephemeris, such that the cross-track uncertainty of predicted stellar occultations is relatively modest by TNO standards.
At the same time, Aya's sustained separation from the Galactic plane implies a low surface density of suitably bright background stars along its apparent path, and hence a very low intrinsic rate of accessible occultation events.
To date, despite the favourable ephemeris, no successful stellar occultation by Aya has been reported.

In this context, we report here on a stellar occultation by Aya that occurred on 2024 May 10 (UT), observed with three facilities in Japan: the Murikabushi telescope at Ishigakijima Astronomical Observatory, the Seimei telescope at Okayama Observatory, and the Kiso Schmidt telescope at Kiso Observatory.
These systems provided high-cadence, high signal-to-noise optical imaging of the target star.
Although a clear occultation by the main body was only detected at a single chord (Ishigaki), the comparatively small ephemeris uncertainties of Aya (Sect.~\ref{sec:obs}) allow us to adopt a Bayesian framework in which the centre of the shadow is treated as a random variable with a prior distribution derived from the orbital covariance.
By combining this ephemeris-informed prior with the detected chord, we are able to place meaningful constraints on the size of Aya's shadow and hence on its equivalent radius.
This strategy follows and extends the single-chord Bayesian occultation analysis demonstrated for asteroid (98943)~Torifune, where physically meaningful constraints were derived from a single detected chord \citep{Arimatsu2024-cc21}.
In addition, the obtained data sets jointly offer an opportunity to constrain any atmosphere or narrow ring system around Aya.
In this paper we present the observations and data reduction in Sect.~\ref{sec:obs}, and describe our results and the constraints on Aya's size, atmosphere, and rings in Sect.~\ref{sec:results}. We also discuss the implications of our results for the physical properties of Aya and for the broader population of mid-sized TNOs in Sect.~\ref{sec:results}, and we summarise our conclusions in Sect.~\ref{sec:concl}.

\section{Observations}
\label{sec:obs}
\begin{figure}[ht!]
   \centering
   \includegraphics[width=0.8\hsize]{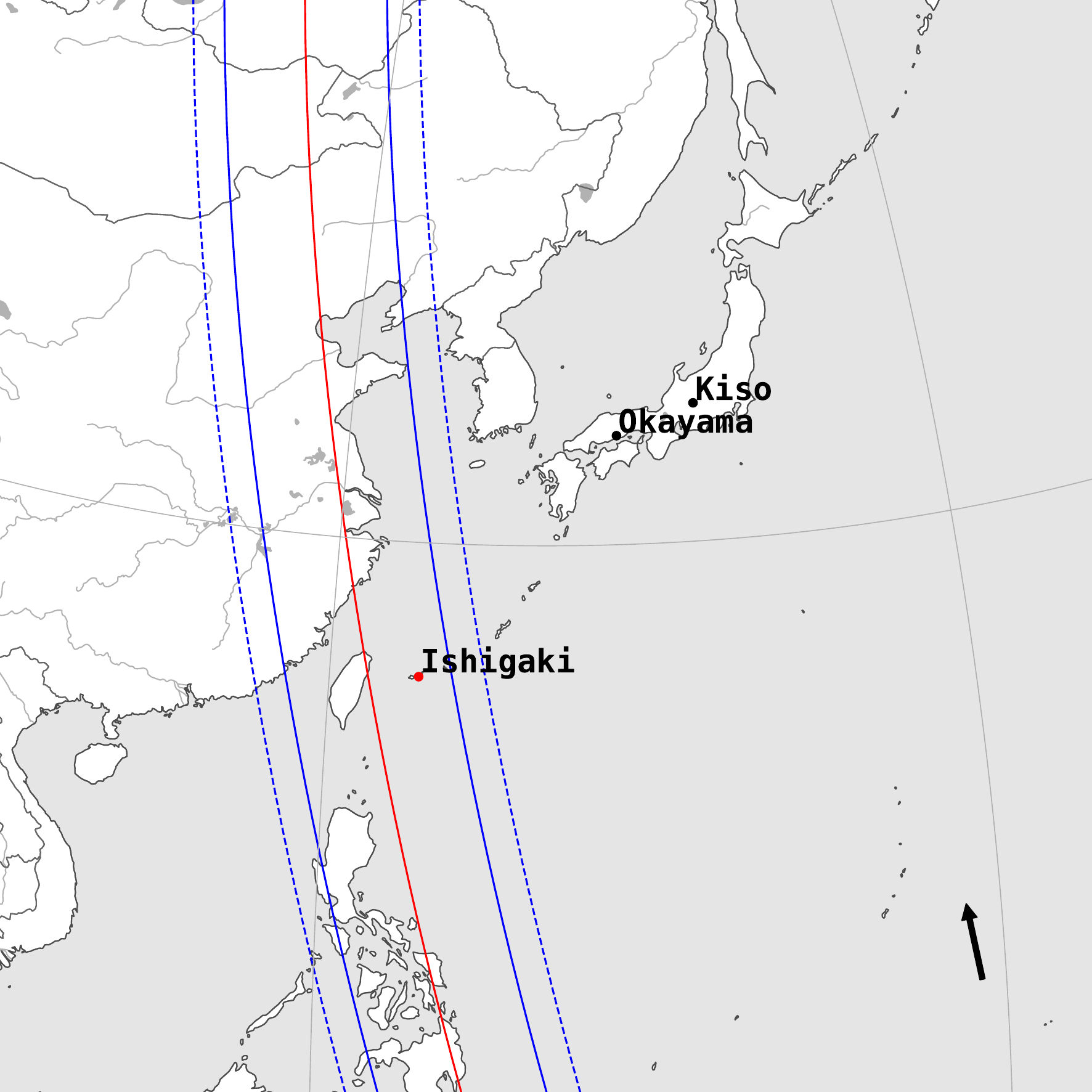}
      \caption{Prediction map of the (55565) Aya stellar occultation event on 10 May 2024. The red line is the central path of the occultation shadow predicted from the JPL34 orbital elements, and the two blue solid lines represent the shadow width of Aya assuming a radius of 384~km. The two blue dashed lines show the $1\sigma$ error range of the predicted shadow path. The red circle indicates the location of the Ishigaki site, where the observed light curve shows a positive detection. The black points show the locations of the negative detections (Okayama and Kiso).}
         \label{fig:predmap}
   \end{figure}

On 2024 May 10, Aya occulted the star Gaia DR3 3773877025854140800 (UCAC4 421-052967) with the Gaia $G$-band magnitude of 13.8 at a velocity relative to the geocentre $v_{\rm rel} = 7.8~{\rm km \, s^{-1}}$.
This event was predicted using an ephemeris for the object (JPL\#34 orbital elements, retrieved from the JPL Horizons system; \citealt{Giorgini1996-horizons}) and the target star from the Gaia DR3 catalogue \citep{Vallenari2023-nk} with the open-source package Stellar Occultation Reduction and Analysis (SORA; \citealt{Gomes-Junior2022-sora}).
Figure~\ref{fig:predmap} shows the prediction map of the occultation event.
The shadow path prediction has a cross-track error of $\sim 3.2$ milliarcseconds,
which corresponds to the projected distance of $\sim 1.0\times 10^2$~km.
The astrometric and photometric properties of the occulted star are summarised in Table~\ref{tab:star}. According to the Gaia DR3 catalogue, the re-normalised unit weight error (RUWE) of the occulted star is 1.17, close to the nominal value (RUWE $\sim 1$), and the duplicity flag is false.
These values indicate that the occulted star is not a double star.

\begin{table}[ht!]
\caption{\label{tab:star}Properties of the occulted star.}
\centering
\small
\begin{tabular}{ll}
\hline\hline
Parameter & Value \\
\hline
Gaia DR3 source ID & 3773877025854140800 \\
RA (ICRS, epoch 2016.0) & 10 07 27.2707 ($\pm0.019$~mas) \\
Dec (ICRS, epoch 2016.0) & $-$05 53 49.148 ($\pm0.018$~mas) \\
$\mu_{\alpha}\cos\delta$~[${\rm mas\,yr^{-1}}$] & $-5.938 \pm 0.023$ \\
$\mu_{\delta}$~[${\rm mas\,yr^{-1}}$] & $-7.850 \pm 0.021$ \\
Parallax~[mas] & $1.066 \pm 0.022$ \\
$G$, $G_{\rm BP}$, $G_{\rm RP}$~[mag] & 13.77, 14.10, 13.28 \\
$g$, $r$, $i$~[mag]\tablefootmark{a} & 14.18, 13.75, 13.62 \\
RUWE & 1.17 \\
Position at event epoch\tablefootmark{b} & 10 07 27.2673, $-$05 53 49.213 \\
Position uncertainty at epoch\tablefootmark{b} & $0.19 \times 0.15$~mas \\
Angular diameter (estimated)\tablefootmark{c} & $\simeq0.01$~mas ($\simeq0.3$~km) \\
\hline
\end{tabular}
\tablefoot{Astrometric and photometric quantities are from Gaia DR3 \citep{Vallenari2023-nk}.\\
\tablefoottext{a}{Synthetic SDSS magnitudes from the Gaia DR3 synthetic photometry catalogue (GSPC).}
\tablefoottext{b}{Astrometric (ICRS) position propagated to the event epoch using the Gaia five-parameter solution and its covariance; the $1\sigma$ uncertainty is dominated by the proper-motion errors over the 8.4-yr baseline.}
\tablefoottext{c}{From $(V-K_{\rm s})$ surface-brightness relations \citep{Kervella2004-sb,vanBelle1999-sb} with $K_{\rm s}=12.31$ from the Two Micron All Sky Survey (2MASS); much smaller than the Fresnel scale ($\simeq1.5$~km).}}
\end{table}

The same occultation event was also predicted by the European Research Council (ERC) Lucky Star\footnote{\url{https://lesia.obspm.fr/lucky-star/occ.php?p=126364}} project, whose predictions rely on the Numerical Integration of the Motion of an Asteroid (NIMA) ephemerides \citep{Desmars2015-nima}.

We carried out observations of the event at three stations in Japan as part of the observation campaign Trans-Neptunian Atmospheres and Belts Analysis through Stellar-occultation Coordinated Observations (TABASCO), part of the Optical and Infrared Synergetic Telescopes for Education and Research (OISTER) programme.
The locations of the observing sites are shown in Fig.~\ref{fig:predmap}, and their observing circumstances are summarised in Table~\ref{tab:sites}.
As indicated in Fig.~\ref{fig:predmap}, the JPL\#34 orbital elements available immediately before the occultation event suggested that an occultation by the main body of Aya would be observable only from Ishigaki. However, we carried out observations at three sites in order to search for possible circum-object material, including rings, with 1--4-m class telescopes.

\begin{table}[ht!]
\caption{\label{tab:sites}Observing circumstances for the 10 May 2024 stellar occultation.}
\centering
\small
\begin{tabular}{lccc}
\hline\hline
Site & Longitude & Aperture  & Observer\\
                 & Latitude  & Camera    &          \\
                 & Altitude  & Exp.\ / cycle time &          \\
                 &           & Time source &          \\
\hline
Ishigaki   & 124\degr 08\arcmin 21.4\arcsec~E & 1.05~m  & H.H.\\
           & 24\degr 22\arcmin 22.3\arcsec~N & MITSuME &    \\
           & 197~m                                & 1.0~s / 3.2~s &    \\
           &                                      & GPS &    \\
\hline
Okayama     & 133\degr 35\arcmin 48.2\arcsec~E & 3.78~m & K.A.\\
            & 34\degr 34\arcmin 36.8\arcsec~N  & TriCCS &  \\
            & 355~m & 0.19~s / 0.19~s  & \\
            &       & GPS & \\
\hline
Kiso        & 137\degr 37\arcmin 31.5\arcsec~E & 1.05~m  & S.T. \\
            & 35\degr 47\arcmin 50.0\arcsec~N & Tomo-e & \\
            & 1,132~m & 0.053~s / 0.053~s & \\
            &         & GPS & \\
\hline
\end{tabular}
\end{table}

At Ishigaki, we conducted observations using the Multicolor Imaging Telescopes for Survey and Monstrous Explosions (MITSuME) camera system mounted on the 1.05-m Murikabushi Telescope at the Ishigakijima Astronomical Observatory, National Astronomical Observatory of Japan.
MITSuME consists of three $1024 \times 1024$ pixel CCD cameras (Apogee Alta U-6), enabling almost simultaneous imaging over a field of view (FoV) of $12.3 \times 12.3~{\rm arcmin^2}$ in three bands ($g'$, $R_c$, and $I_c$), with effective wavelengths of 477, 649, and 802~nm, respectively.
In the present observation, we obtained sequential images with an exposure time of $1.0$~s per frame.
Since the overhead time of MITSuME is approximately 2.2~s, the total time required for a single frame was 3.2~s, corresponding to a frame rate of 0.31~fps.
At Okayama, we used the Tricolor CMOS Camera and Spectrograph (TriCCS) camera system mounted on the Seimei telescope at Okayama Observatory.
The TriCCS consists of three CMOS sensors offering high-cadence (up to $\sim 100$ frames per second, fps) three-band imaging over a $12.6 \times 7.5~{\rm arcmin^2}$ FoV.
In the present observation, we monitored the occulted star simultaneously in the $g$, $r$, and $i$ bands, whose central wavelengths are approximately 481, 617, and 752~nm, respectively. The exposure time was $0.19$~s per frame, corresponding to a frame rate of $\sim 5.2$~fps.
At Kiso, no-filter images were obtained using the Tomo-e Gozen camera mounted on the 1.05 m f/3.1 Schmidt camera at Kiso Observatory.
In the present observation, we used a single CMOS sensor with a fast-readout mode offering a region-of-interest FoV of $7.9 \times 4.8~{\rm arcmin^2}$.
The exposure time for each frame was $0.053$ s, corresponding to the frame rate of $\sim 18.9$ fps.

Observations were conducted during twilight at all three observing sites (Ishigaki, Okayama, and Kiso), and time-series images were successfully acquired.
Standard aperture photometry of the occulted star was performed for each data set.
To correct possible flux fluctuations due to rapid changes in atmospheric conditions,
we carried out differential photometry using non-occulted stars detected simultaneously on the same sensor. At Ishigaki, two such stars were available in each of the three bands, and were combined with signal-to-noise weights to form the reference flux by which the flux of the occulted star was divided. Their mutual flux ratio provides a control light curve that is unaffected by this normalisation and is used in Appendix~\ref{app:control} to verify the stability of the photometry.

\section{Results and discussion}
\label{sec:results}

\subsection{Main body occultation}
\label{sec:mainbody}
\begin{figure}[ht!]
  \centering
  \includegraphics[width=0.85\hsize]{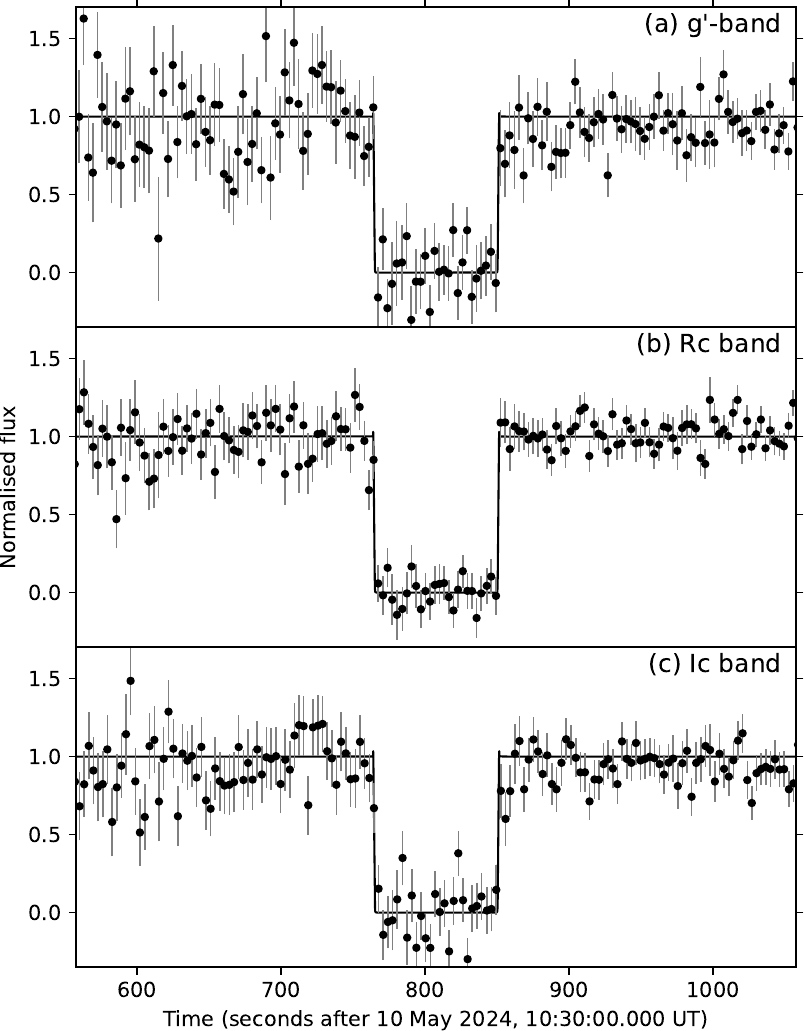}
    \caption{Stellar occultation light curves obtained at the Ishigaki station using the 1.05-m Murikabushi Telescope/MITSuME: (a) $g'$, (b) $R_c$, and (c) $I_c$ bands.
    The points with error bars are observed photometric points, whereas the solid lines represent the best-fit synthetic light curves from the joint three-band fit.
    }
    \label{fig:LC_Ishigaki}
\end{figure}

\begin{figure}[ht!]
  \centering
  \includegraphics[width=0.85\hsize]{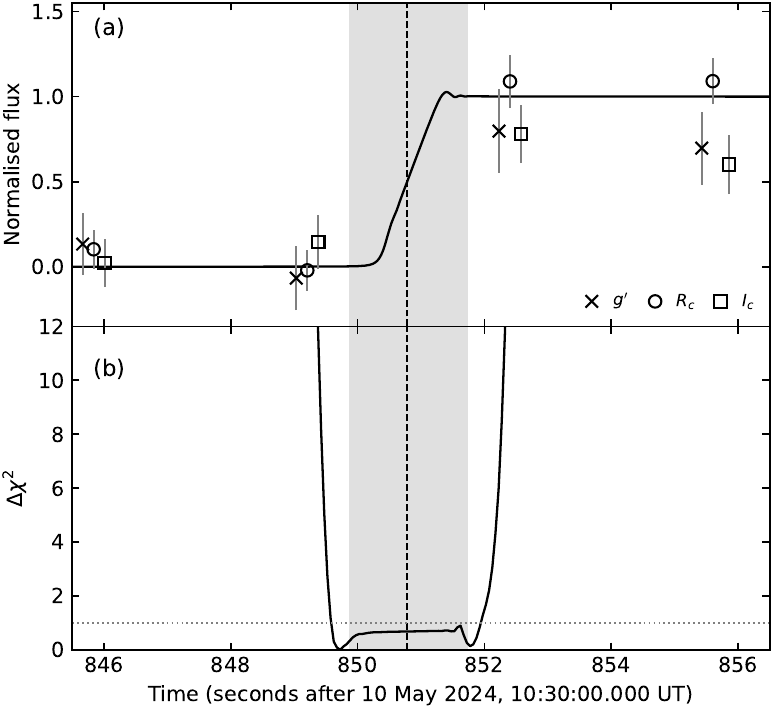}
    \caption{Diagnostic of the emersion timing.
    (a) Light curve around the emersion in the three bands ($g'$: crosses, $R_c$: open circles, $I_c$: open squares),
    with the best-fit joint model for the $g'$ band (convolved with the 1~s exposure) overlaid as the solid black line.
    The models for the other two bands are nearly identical because the finite-exposure smearing dominates over the wavelength-dependent diffraction feature.
    (b) $\Delta\chi^2$ of the joint three-band fit as a function of the emersion time. In both panels the dashed vertical line marks the adopted emersion time (10h~44m~10.78s) and the shaded band the 1.85~s interval that is covered by none of the three cameras.
    The dotted horizontal line marks $\Delta\chi^2=1$, corresponding to the $1\sigma$ level.
    The $\Delta\chi^2$ curve is nearly flat within this interval. Therefore, the emersion time cannot be constrained more precisely.
    The two shallow minima at its ends differ by only $\Delta\chi^2=0.14$ and are not statistically significant.
    We therefore adopt the posterior mean rather than either minimum.}
    \label{fig:egress}
\end{figure}

\begin{figure}[ht!]
  \centering
  \includegraphics[width=0.8\hsize]{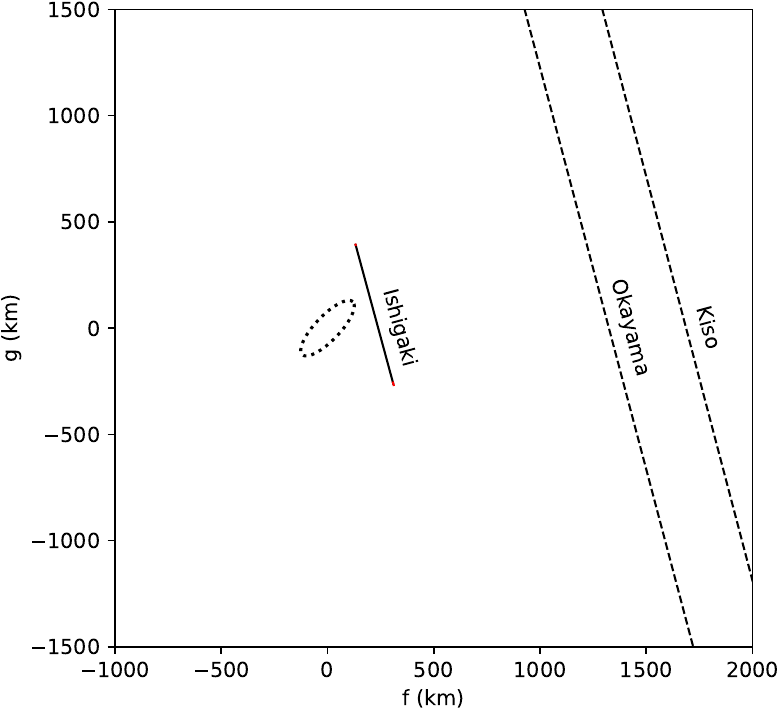}
    \caption{Occultation chords projected onto the sky plane as seen from three observation sites.
The $f$ and $g$ axes indicate the east-west and north-south directions in the plane, respectively.
The solid black line represents the positive occultation chord observed at the Ishigaki station.
The red segments are the uncertainties ($1\sigma$ level) at the extremities of the positive chord.
The two dashed lines represent the negative chords at the Okayama and Kiso stations.
The dotted ellipse corresponds to the $1\sigma$ constraint on the shadow centre of Aya, which combines the ephemeris uncertainty from the JPL39 orbital elements and the astrometric uncertainty of the occulted star from the Gaia DR3 catalogue.
    }
    \label{fig:chord}
\end{figure}

Out of the three stations, a single positive detection was obtained at Ishigaki, which is consistent with the shadow path prediction (see Fig.~\ref{fig:predmap}).
Figure~\ref{fig:LC_Ishigaki} shows the light curves of the occultation obtained at the Ishigaki station (the 1.05-m Murikabushi Telescope at the Ishigakijima Astronomical Observatory).
We obtained the positive-detection light curves in three bands ($g'$, $R_c$, and $I_c$) simultaneously.
The apparent immersion and emersion times of the occultation were measured by fitting a Fresnel-diffraction model for a semi-infinite straight edge convolved by the apparent angular diameter of the occulted star to the observed light curve.
The model then integrates the monochromatic diffraction pattern over wavelength, weighted by the spectral response of each band.
We performed a single joint fit to the three band light curves ($g'$, $R_c$, and $I_c$) simultaneously.
The immersion and emersion times are shared across the three bands, since the geometric event is common to all wavelengths, while the wavelength-dependent diffraction profile is computed for each band.
For the fitting,
we determined the baseline and zero-flux levels for each band from the data points taken 16--60~s outside the event and more than 6~s inside the occultation, respectively.
Their uncertainties were then propagated to the scaled flux errors.
Consequently, the model requires only two free parameters, the immersion and emersion times, or equivalently,
the mid-time and duration of the event.
The best-fit times are those that minimise the total $\chi^2$ summed over the three bands.
For each band,
the fit uses 12 samples within 20~s of immersion and another 12 within 20~s of emersion,
yielding 24 samples per band and 72 in total,
with 70 degrees of freedom.
The minimum total $\chi^2$ is 56.2, corresponding to a reduced $\chi^2$ of 0.80.
The timing uncertainties are derived from the variation of the total $\chi^2$ about this minimum, as described below.
This joint approach exploits the simultaneity of the three-band photometry and yields tighter and mutually consistent timing constraints than separate single-band fits.
Table~\ref{tab:times} lists the times obtained from independent fits to each band alongside the joint result, together with the number of samples and degrees of freedom of each fit.
The best-fit synthetic profiles are shown as solid lines in Fig.~\ref{fig:LC_Ishigaki}.
The immersion and emersion times are estimated to be 10h 42m $44.59^{+0.11}_{-0.09}$s~UT and 10h 44m $10.78\pm0.74$s~UT, respectively.
The immersion and emersion times were determined differently.
All three cameras used 1.00~s exposures on a 3.20~s cycle, with exposure start times offset by 0.17 and 0.26~s.
At immersion, the star remained visible for approximately 90\% and 64\% of the $R_c$ and $I_c$ exposure times, respectively, yielding two partially occulted measurements.
These measurements yield a well-defined minimum in the $\chi^2$ profile, and we report the corresponding $\Delta\chi^2<1$ interval.
A solution with immersion during the subsequent dead time is disfavoured by $\Delta\chi^2\simeq5.8$.
At the emersion, the star reappeared within a 1.85~s interval (10h~44m~09.88s to 11.73s) covered by none of the three cameras (Fig.~\ref{fig:egress}a).
All emersion times within this interval yield identical model fluxes, resulting in a flat $\chi^2$ profile (Fig.~\ref{fig:egress}b).
The two shallow minima at the interval boundaries differ by only $\Delta\chi^2=0.14$ and each arise from a single sample at the $1\sigma$ level.
Neither minimum is statistically significant.
We therefore adopt the mean and standard deviation of the posterior probability density,
$P(t_{\rm e})\propto\exp[-\chi^2(t_{\rm e})/2]$,
derived from the $\chi^2$ summed over the three bands under a uniform prior (Fig.~\ref{fig:egress}b): $10\mathrm{h}\,44\mathrm{m}\,10.78\mathrm{s}\pm0.74$~s.
The quoted $1\sigma$ uncertainty is thus derived from the joint $\chi^2$ profile.
For this flat profile,
the posterior standard deviation avoids the ambiguity of referencing a $\Delta\chi^2=1$ interval to either shallow minimum.
This value agrees to within 0.03~s with the centre and half-width of the uncovered interval ($10.81\pm0.93$~s), which is independent of the adopted model.
The obtained chord duration of the occultation is $86.19\pm0.75$~s.
Figure~\ref{fig:chord} shows the positive occultation chord projected onto the sky plane.
The observed positive chord length is $678 \pm 6$~km.

A number of samples in Fig.~\ref{fig:LC_Ishigaki} lie below the normalised flux level of unity outside the occultation, most noticeably in the $g'$ and $I_c$ bands. These excursions are consistent with photometric noise.
The dispersion of the out-of-occultation samples (21, 13, and 15\% in $g'$, $R_c$, and $I_c$, respectively) is fully accounted for by the estimated per-sample uncertainties, whose median values are 20.2, 13.5, and 16.5\%.
No sample departs from unity by more than $2.8\sigma$ of its own uncertainty in any band, and the numbers of samples deviating by more than $2\sigma$ (6, 5, and 6 out of 136) agree with the Gaussian expectation of 6.2.
Since the three bands were recorded simultaneously, a genuine extinction event would appear in all of them at the same epoch.
None of the samples lying more than twice the out-of-occultation dispersion below unity (2 in $g'$ and 9 in $I_c$), however, coincides between the two bands to within 1~s.
Finally, none of these features appear in the comparison stars (Appendix~\ref{app:control}),
confirming that the excursions are not residual instrumental or atmospheric variations left after differential photometry.
Instead, they are consistent with random photon and detector noise in the individual measurements of the occulted star,
which explains both their amplitude and frequency as calculated above.

\begin{table}[ht!]
\caption{\label{tab:times}Immersion and emersion times from the single-band and joint fits.}
\centering
\small
\renewcommand{\arraystretch}{1.3}
\begin{tabular}{lcc}
\hline\hline
Fit & Immersion & Emersion \\
    & 10h 42m [s] & 10h 44m [s] \\
\hline
$g'$   & $44.66^{+2.26}_{-0.27}$ & $11.92^{+0.25}_{-2.46}$ \\
$R_c$  & $44.56^{+0.14}_{-0.13}$ & $11.75^{+0.21}_{-2.16}$ \\
$I_c$  & $44.66^{+0.14}_{-0.12}$ & $12.24^{+0.12}_{-2.62}$ \\
Joint  & $44.59^{+0.11}_{-0.09}$ & $10.78\pm0.74$ \\
\hline
\end{tabular}
\tablefoot{Times are UT on 2024 May 10. For the single-band fits and for the joint immersion the quoted value is the profile $\chi^2$ minimum with its $\Delta\chi^2<1$ interval, where $\Delta\chi^2$ refers to the total $\chi^2$ (22 and 70 degrees of freedom for the single-band and joint fits, respectively). The joint emersion corresponds to the posterior mean and standard deviation derived from the summed $\chi^2$ profile, because the profile $\chi^2$ is flat over the interval covered by none of the cameras (Sect.~\ref{sec:mainbody}). The single-band emersion minima listed here all fall at the late end of that interval, but the differences with respect to its early end are only $\Delta\chi^2 = 0.69$, $0.18$, and $0.70$ for $g'$, $R_c$, and $I_c$, and are therefore not significant. The large upper uncertainty of the $g'$ immersion reflects the fact that, of the three staggered exposures, only those of $R_c$ and $I_c$ straddled the stellar disappearance.}
\end{table}

\subsection{MCMC fit}
\label{sec:mcmc}
In the following, we estimate the equivalent size of the occulting shadow of Aya
by considering the single detected chord of the occultation and the ephemeris.
We assume a circular projected shadow with radius $r$.
The derived radius should therefore be understood as an equivalent radius of the projected shadow, not as a determination of the actual shape of the body (see Sect.~\ref{sec:size}).
To estimate the posterior probability distribution of the size and shape parameters in the Bayesian approach,
we employ a Markov chain Monte Carlo (MCMC) scheme.
We perform fits of a circular shadow to the observed chord extremities using the open source ensemble sampler code \texttt{emcee} \citep{Foreman-Mackey2013-ij}.
A circular shadow centred at $(f_c, g_c)$ with radius $r$ is fitted to the data points of the observed extremities by
\begin{equation}
    \label{eq1}
    \chi^2=\sum_i^N \frac{\left(r_{i,\mathrm{obs}}-r\right)^2}{\sigma_{r, i}^2},
\end{equation}
where $(r_{i,\mathrm{obs}}-r)$ is the difference between the distance of observed chord extremity $i$ from the model shadow centre, $r_{i,\mathrm{obs}}$, and $r$, and $\sigma_{r,i}$ is the $1\sigma$ uncertainty in the radial direction.
We parametrise the model by the shadow centre coordinates $(f_c,g_c)$ and the shadow radius $r$.
We run 32 chains of MCMC walkers with randomised initial parameters.
For each chain we run 30000 steps with a burn-in of the first 5000 steps.
To verify convergence, we repeated the entire procedure using six independent sets of random seeds.
The resulting median parameters differ by only $\sim 1$~km in $r$ and $\sim 2$~km in $(f_c, g_c)$, well below the statistical uncertainties reported below.
The values reported in this paper are derived from the six chains combined.

We impose an informative prior on the shadow centre $(f_c,g_c)$ based on the predicted uncertainty of the occultation geometry.
The prior is modelled as a two-dimensional Gaussian distribution whose covariance matrix represents the combined uncertainty of the latest JPL\#39 ephemeris of Aya available as of March 2026 and the astrometric uncertainty of the occulted star taken from the Gaia DR3 catalogue \citep{Vallenari2023-nk}.
The resulting $1\sigma$ error ellipse in the $f$-$g$ plane has major and minor axes of $172$~km and $58$~km, respectively,
with a position angle of ${\rm PA}=136^{\circ}$ (see Fig.~\ref{fig:chord}).
We assume a uniform prior for the shadow radius $r$ over $r > 0$.
The posterior probability is then sampled, using an ensemble of walkers initialised near the prior mean of $(f_c,g_c)$ and a characteristic radius estimated from the observed chord geometry.
From the converged MCMC chain we derive the marginalised posterior distributions of $f_c$, $g_c$, and $r$.
The sensitivity of the derived size to these prior assumptions is examined in Sect.~\ref{sec:robust}.

\subsection{Constraints on the projected size of Aya}
\label{sec:size}
\begin{figure}[ht!]
  \centering
  \includegraphics[width=0.85\hsize]{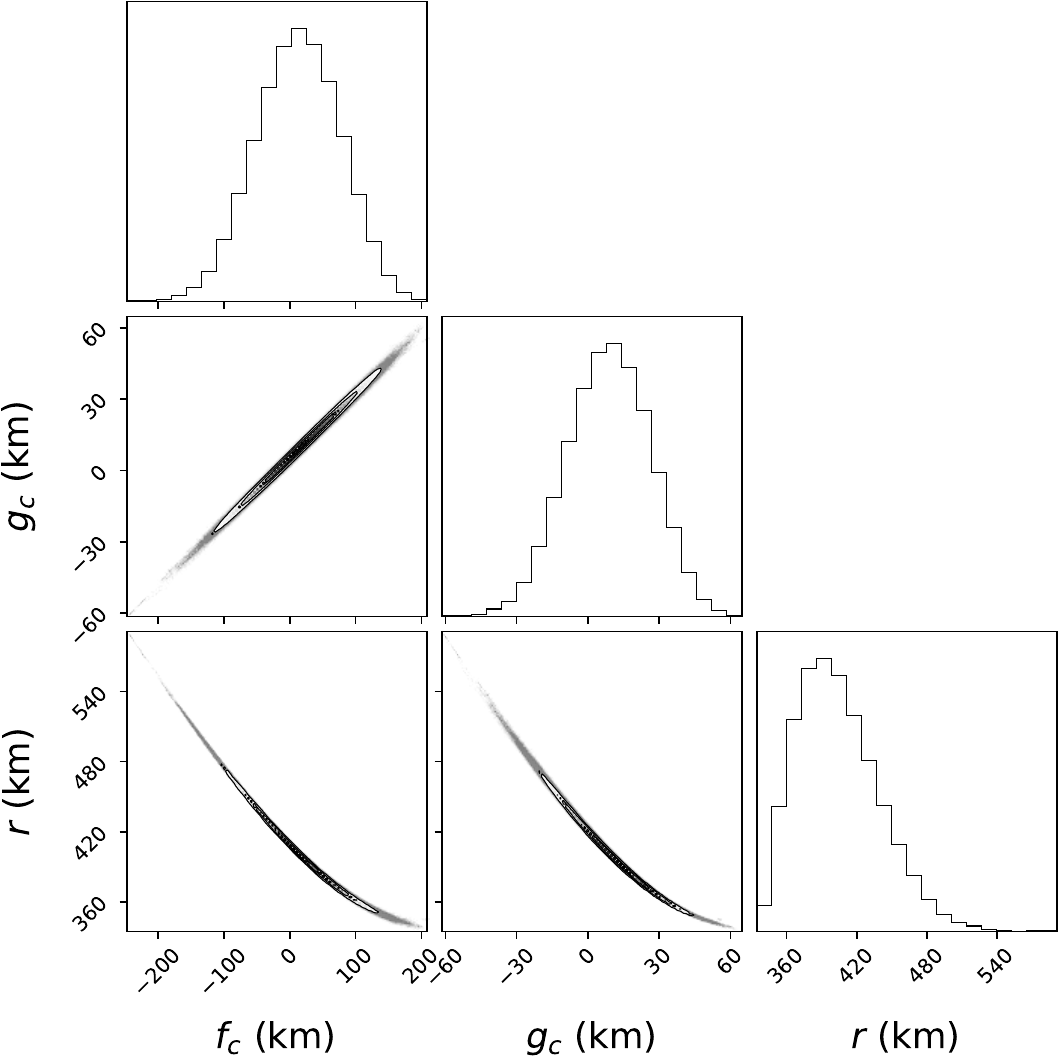}
    \caption{Posterior distributions of the parameters ($f_c$, $g_c$, and $r$) derived from the circular shadow fit.
    }
    \label{fig:posterior}
\end{figure}

Figure~\ref{fig:posterior} shows the posterior distributions of the parameters obtained from the MCMC fit, and the resulting best-fit parameters with their uncertainties are listed in Table~\ref{tab:best-fit}.
A single chord constrains the shadow geometry mainly along the direction of the chord, while leaving a strong degeneracy between the shadow radius and the location of the shadow centre perpendicular to the chord.
As a result, the posterior distributions exhibit noticeable skewness.
Nevertheless, the equivalent shadow radius is meaningfully constrained by the data themselves, without imposing a strong prior on the radius.

\begin{table}[H]
\caption{\label{tab:best-fit}Best-fit parameters obtained from the MCMC fit.}
\centering
\small
\renewcommand{\arraystretch}{1.3}
\begin{tabular}{lc}
\hline\hline
Parameter & Best-fit values\\
\hline
$f_c$~[km] & $14^{+67}_{-62}$ \\
$g_c$~[km] & $9^{+18}_{-17}$ \\
$r$~[km]   & $401^{+28}_{-40}$ \\
\hline
\end{tabular}
\end{table}

\begin{figure}[ht!]
  \centering
  \includegraphics[width=0.8\hsize]{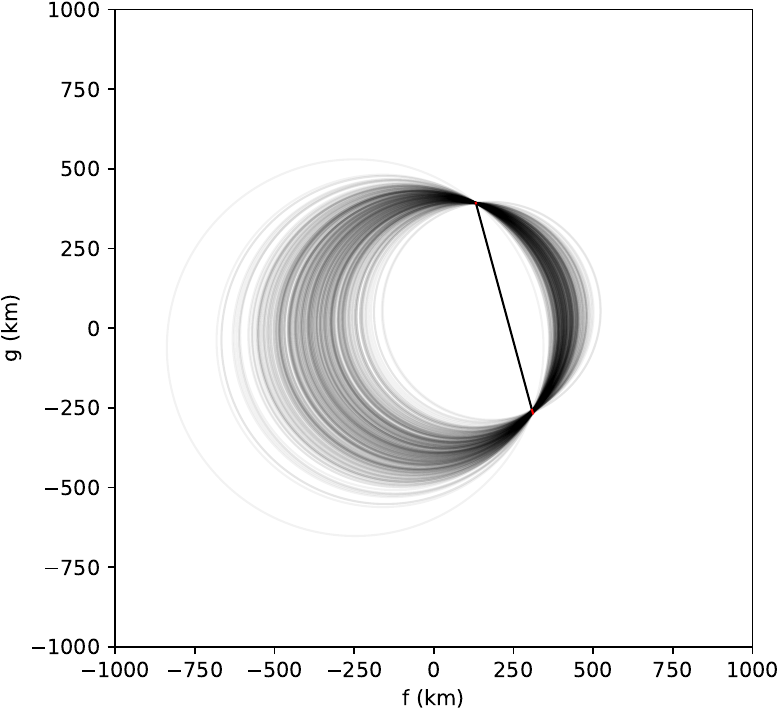}
    \caption{Positive occultation chord obtained at the Ishigaki station, projected on the sky plane (the same as in Fig.~\ref{fig:chord}), overlaid with sample circular occultation shadows of Aya drawn from the posterior distribution of the MCMC fit.
    }
    \label{fig:chord_fit}
\end{figure}

Our MCMC analysis yields an equivalent shadow radius of $r = 401^{+28}_{-40}$~km.
The quoted uncertainties correspond to the 68\% highest-posterior-density intervals of the marginalised posterior distributions.
Figure~\ref{fig:chord_fit} shows circular shadows drawn from the posterior distribution overlaid on the observed chords.

We emphasise that this value is the radius of the circle fitted to the projected shadow, i.e., an estimate of the equivalent radius of Aya at the epoch of the event, and not a measurement of the actual, possibly aspherical, shape of the object.
A single chord cannot constrain the projected elongation or orientation of the limb.
The low amplitude of the rotational light curve of Aya \citep{Ortiz2006-dl,Thirouin2010-lj} suggests a nearly circular projected profile, supporting the adopted circular-shadow model, but a residual shape-dependent bias of the order of the projected elongation cannot be excluded.
We derived the geometric albedo at V-band from the following equation proposed by \citet{Sicardy2011-dl}:
\begin{equation}
    p_V=10^{0.4\left(V_{\rm \odot}-H_V\right)} /(r/1~{\rm au})^2,
\end{equation}
where $V_{\rm \odot}$ is the V-band magnitude of the Sun at 1~au, i.e., $V_{\rm \odot} = -26.76$ \citep{Willmer2018-sun}.
Assuming the $H_V = 3.593 \pm 0.023$ obtained by the previous study \citep{Alvarez-Candal2016-sf},
the geometric albedo is estimated to be $p_V = 0.101^{+0.023}_{-0.013}$.
This size and albedo estimate can be directly compared with previously published radiometric sizes for Aya.
Early Spitzer/MIPS thermal modelling by \citet{Cruikshank2005-ii} reported $r = 350\pm 25$~km with a relatively high geometric albedo $p_V \sim  0.17\pm0.03$.
A more precise constraint was obtained by combining Herschel/PACS and Spitzer/MIPS measurements, yielding $r = 384\pm 19$~km and $p_V = 0.112^{+0.012}_{-0.011}$.
Our occultation-based equivalent radius is therefore larger by $\sim 4\%$, and the corresponding geometric albedo lower by $\sim 10\%$, than the Herschel+Spitzer best values, although both remain statistically consistent within the respective uncertainties. Recent studies have suggested that radiometric sizes derived from thermal infrared emission may be underestimated by about 6\% on average relative to occultation-based sizes \citep{Sicardy2024-fu}. The present result is apparently consistent with this trend.
We note that the two size estimates are not strictly equivalent
because radiometry averages the emitting cross-section over multiple observations,
whereas the occultation measures the instantaneous projected limb.
Aya's rotational light-curve amplitude ($\lesssim0.1$~mag; \citealt{Ortiz2006-dl,Thirouin2010-lj}) implies variations of $\lesssim5\%$ in equivalent radius, comparable to the 4\% difference found above.
We also note that the Herschel+Spitzer solution implies an $H_V$ within $\sim0.02$~mag of the value adopted here, so the albedo comparison is made on a consistent photometric scale.
The higher early Spitzer albedo of \citet{Cruikshank2005-ii} largely reflects its brighter $H_V \simeq 3.3$.

\subsection{Robustness of the size estimate}
\label{sec:robust}

Since only one positive chord was obtained, the inferred size depends partly on the prior for the shadow centre and therefore on the adopted ephemeris.
We assessed this dependence using two tests.

First, we repeated the fit with a different ephemeris.
The chord positions in the $f$--$g$ plane are defined relative to the predicted shadow centre, so the ephemeris affects the fit only through the centre of the Gaussian prior.
We replaced the JPL\#39 solution with JPL\#34, which was available immediately before the event (Sect.~\ref{sec:obs}).
The JPL\#39 solution includes about two additional years of astrometric data.
At the event epoch, the shadow centres predicted by the two solutions differ by only $\simeq 12$~km ($\simeq 0.4$~mas).
This offset is more than an order of magnitude smaller than the dimensions of the $1\sigma$ prior ellipse ($172~{\rm km} \times 58$~km).
With JPL\#34, the posterior centre shifts by the same offset, but the equivalent radius changes by less than 1~km.
This change is negligible compared with the statistical uncertainty.
The inferred size is therefore insensitive to the adopted ephemeris solution.

Second, we estimated the size without assuming a position for the shadow centre, following the single-chord Bayesian framework proposed by \citet{Morgado2026-single}.
For a circular shadow of radius $r$, this method assumes that the chord impact parameter is uniformly distributed over $(-r,r)$.
The likelihood of a chord of length $L$ is $P(L|r) = L/(4r\sqrt{r^{2}-(L/2)^{2}})$ for $0<L<2r$, convolved with the chord-length uncertainty.
We adopted a collisional size-distribution prior $P(r) \propto r^{-q}$ with $q=2.5$.
For our chord ($L = 678 \pm 6$~km), the posterior mode is $r = 341$~km, close to the geometric limit of $L/2 = 339$~km.
The 68\% and 95\% highest-posterior-density credible intervals are 334--400~km and 331--632~km, respectively.
The posterior mode changes by less than 1~km when the prior slope is varied over $q = 1.7$--$8.2$, the range measured for the trans-Neptunian size distribution \citep{Fraser2014-Hdist,Petit2023-ossos}.
Our ephemeris-informed estimate of $r = 401^{+28}_{-40}$~km lies near the upper edge of the 68\% interval, consistent with a non-diametral chord.

Single- and few-chord occultations have been used before to constrain TNO and Centaur sizes \citep{Rommel2020-astrom}.
A relevant example is (119951) 2002~KX$_{14}$, for which a single $415 \pm 1$~km chord yielded an equivalent diameter of $365^{+30}_{-21}$~km \citep{Alvarez-Candal2014-kx14}, whereas subsequent multichord observations revealed a markedly elliptical projected limb ($241 \times 157$~km) with an area-equivalent diameter of $389.2 \pm 8.7$~km \citep{Rizos2025-kx14}.
In that case the single chord happened to sample the body close to its projected long axis.
This underlines the distinction, made throughout this paper, between the equivalent radius derived here and the actual, unknown projected shape of Aya.

\subsection{Constraints on a possible atmosphere}

\begin{figure}[ht!]
  \centering
  \includegraphics[width=0.85\hsize]{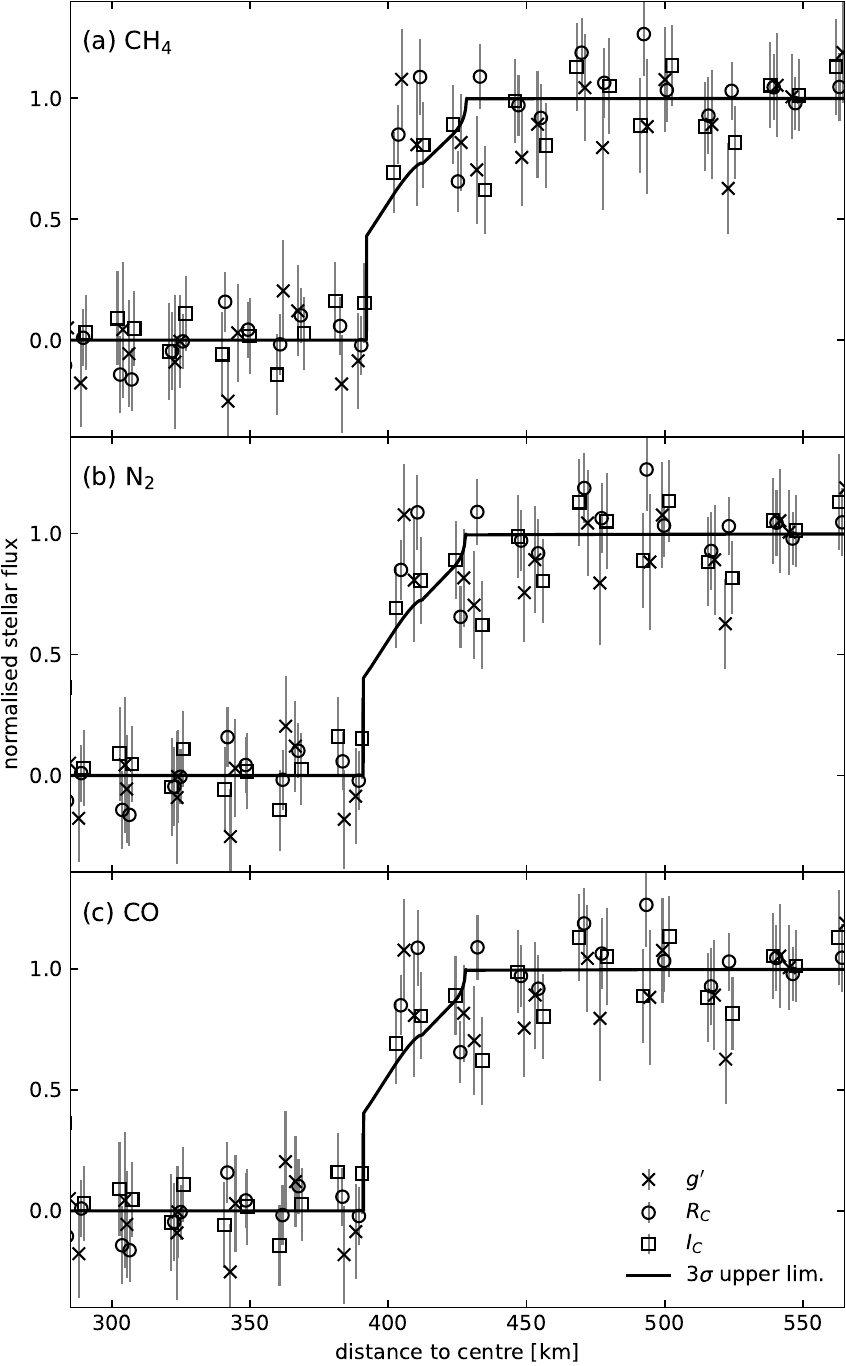}
    \caption{
Observed radial light-curve profile as a function of distance to the shadow centre. The same observed fluxes are shown in the three panels, each compared with the refraction model for one assumed composition, scaled to its 3$\sigma$ surface-pressure upper limit: (a) CH$_4$, $p_0=60$~nbar; (b) N$_2$, $p_0=90$~nbar; (c) CO, $p_0=80$~nbar. Crosses, open circles and squares with error bars show the normalised fluxes in the $g'$, $R_c$, and $I_c$ bands, respectively. All three models assume the same temperature profile, rising from $T=44$~K at the surface to an isothermal $T=106$~K above $\sim$20~km. }
    \label{fig:atmosphere}
\end{figure}

To place observational constraints on any possible surface atmosphere, we applied atmospheric refraction models to the simultaneous three-band photometry ($g'$, $R_c$, $I_c$) obtained with Murikabushi/MITSuME at the Ishigaki station, following previous atmospheric-constraint studies \citep{Arimatsu2019-sk,Santos-Sanz2022-huya,Proudfoot2025-quaoar,Benedetti-Rossi2019-vs2}.
This analysis is particularly motivated by the recent occultation-based report of a possible atmosphere around the plutino (612533)~2002~XV\textsubscript{93}, a mid-sized TNO in a radius regime broadly comparable to Aya, for which the observed light curves were interpreted as refractive attenuation by a tenuous atmosphere with a surface pressure of order 100--200~nbar \citep{Arimatsu2026NatAstro}.
We therefore tested whether Aya could host a similar atmosphere.
Since Aya's surface composition is only weakly constrained observationally \citep{Vilenius2014-wt,Cruikshank2005-ii}, we considered the three principal super-volatiles expected on cold TNO surfaces: CH$_4$, N$_2$, and CO.
For each gas we assumed a pure atmosphere in hydrostatic equilibrium, with the corresponding molecular refractivity and molecular mass, and a common Pluto-like temperature profile
rising from $T = 44~{\rm K}$ at the surface (corresponding to the average surface temperature derived from mid- to far-infrared observations by \citealt{Vilenius2014-wt}) to an isothermal $T = 106~{\rm K}$ above $\sim 20$ km, as expected when the stratosphere is warmed by absorption of solar near-infrared radiation by (trace) CH$_4$ molecules.
We computed refractive attenuation via ray tracing and compared the predicted transmission as a function of distance from the shadow centre with the observed fluxes.
 We treated the surface pressure $p_0$ as a free parameter and defined the 3$\sigma$ upper limit as the maximum pressure satisfying $\chi^2 < \chi^2_{\rm min} + 9$.
 We obtain 3$\sigma$ upper limits of $p_0 = 60$, $90$, and $80$~nbar for CH$_4$, N$_2$, and CO, respectively.
 The three best-fit models are compared with the observed fluxes in Fig.~\ref{fig:atmosphere}.

\begin{figure*}[t]
  \centering
  \includegraphics[width=0.85\textwidth]{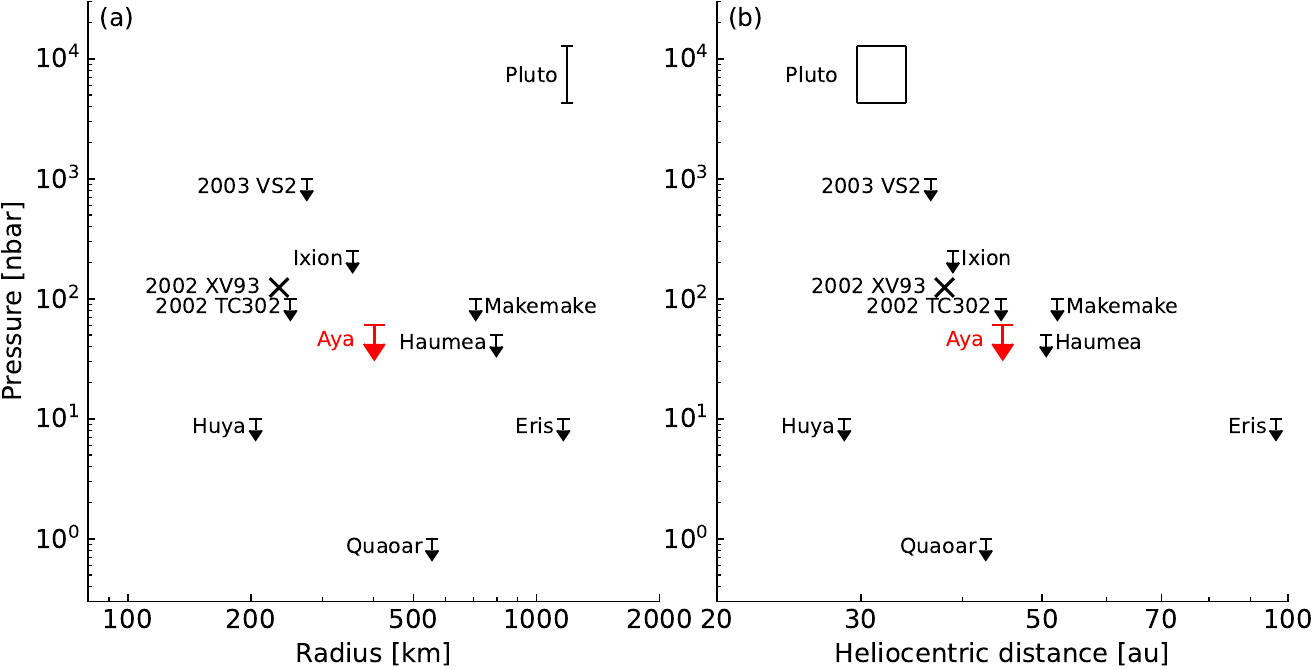}
    \caption{
Observational constraints on the atmospheric surface pressure of Aya as a function of (a) TNO radius and (b) heliocentric distance at the time of observations.
The red arrows are the $3\sigma$ upper limit derived from this study for the CH$_4$ case ($p_0=60$~nbar). The black line and box show the range of surface pressures on Pluto between 1988 and 2020 \citep{Sicardy2003-pluto,Meza2019-pluto,Arimatsu2020-wv,Yuan2023-pluto}. The crosses show the best-fit surface pressure constraints for 2002 XV\textsubscript{93} in cases with a CH\textsubscript{4} atmosphere \citep{Arimatsu2026NatAstro}. The black arrows represent $3\sigma$ upper limits from previous stellar occultation studies for the CH\textsubscript{4} atmosphere case \citep{Santos-Sanz2022-huya,Ortiz2017-mw,Proudfoot2025-quaoar,Sicardy2011-dl,Ortiz2012-sc,Levine2021-ixion,Benedetti-Rossi2019-vs2,Ortiz2020-tc302}.}
    \label{fig:atm_comparison}
\end{figure*}

Figure~\ref{fig:atm_comparison} compares the atmospheric-pressure constraint derived for Aya with previous occultation-based constraints as a function of object radius (panel a) and heliocentric distance at the observing epoch (panel b). Although many TNOs are now known, robust atmospheric constraints remain sparse in the $R\sim200$--$500$~km regime \citep{Santos-Sanz2022-huya,Benedetti-Rossi2019-vs2,Levine2021-ixion,Ortiz2020-tc302}. Our result therefore adds a new and meaningful constraint in this size range.
In particular, Aya now provides an important comparison point to 2002~XV\textsubscript{93} in the same mid-sized-TNO regime.
Our upper limit for Aya ($p_0<60$~nbar at 3$\sigma$ for the CH$_4$ composition) suggests that such atmospheres are not ubiquitous among objects of this class. This trend is consistent with the scenario discussed by \citet{Arimatsu2026NatAstro}, in which these atmospheres may be transient and driven by short-timescale processes rather than being a long-lived, common state.

\subsection{Constraints on ring and other circumplanetary extinction}

To constrain narrow rings and other localised circumplanetary extinction,
we searched all three light curves for short-duration extinction events, following the ring-search strategy adopted in recent stellar-occultation analyses \citep{Santos-Sanz2022-huya,Rommel2025-huya,BragaRibas2026-quaoar}.

For each station we used the band giving the highest photometric precision: the $r$ band of Seimei/TriCCS at Okayama; the $R_c$ band of Murikabushi/MITSuME at Ishigaki; and the unfiltered Tomo-e Gozen frames at Kiso.
At Ishigaki, which recorded the positive chord, the interval during which the star was hidden by the body itself was excluded.

We constructed two time series from each data set: (i) unbinned single-exposure fluxes and (ii) fluxes binned by a factor of 10. For the Okayama $r$-band photometry their effective temporal resolutions are 0.193~s and 1.93~s, respectively.
Using the shadow velocity at that station, $v_{\rm rel}=7.86~{\rm km\,s^{-1}}$, these correspond to projected spatial scales of 1.52~km and 15.2~km along the chord direction.
The equivalent scales are 7.9 and 251~km at Ishigaki and 0.41 and 4.2~km at Kiso.
All lengths and distances quoted in this ring analysis are measured in the sky plane, i.e., in the plane perpendicular to the line of sight.
Since the orientation of Aya's rotational pole is unknown, they cannot be deprojected onto the equatorial plane of the body, where rings are expected to lie.
The sampling and detection parameters of the three stations are collected in Table~\ref{tab:ring}.

\begin{table}[ht!]
\caption{\label{tab:ring}Ring-search parameters and limits for the three stations.}
\centering
\small
\begin{tabular}{lccc}
\hline\hline
                                  & Ishigaki & Okayama & Kiso \\
                                  & MITSuME  & TriCCS  & Tomo-e \\
\hline
Band                              & $R_c$   & $r$     & clear  \\
$t_{\rm exp}$~[s]                 & 1.0     & 0.193   & 0.053  \\
Cycle time~[s]                  & 3.2     & 0.193   & 0.053  \\
$\Delta$~[km]\tablefootmark{a}    & 7.9     & 1.52    & 0.41   \\
$\sigma_\phi$                     & 0.126   & 0.040   & 0.071  \\
$p'_{\rm min}$\tablefootmark{b}   & 0.38    & 0.12    & 0.21   \\
$p'_{\rm min}$\tablefootmark{c}   & 0.17    & 0.038   & 0.070  \\
Coverage~[km]\tablefootmark{d}    & 430--2130 & 1260--5580 & 1610--8060 \\
Coverage~[$r$]\tablefootmark{d}     & 1.1--5.4 & 3.2--14 & 4.1--20 \\
\hline
\end{tabular}
\tablefoot{
\tablefoottext{a}{Projected sampling length of a single frame, $v_{\rm rel}t_{\rm exp}$.}
\tablefoottext{b}{Unbinned series, $k=3$.}
\tablefoottext{c}{Series binned by a factor of 10, $k=3$; the corresponding sampling lengths are 251, 15.2 and 4.2~km.}
\tablefoottext{d}{Range of projected (sky-plane) distance from the fitted shadow centre actually sampled, in kilometres and in units of the equivalent radius $r=401$~km derived in Sect.~\ref{sec:size}.}
}
\end{table}

We define the apparent opacity as
\begin{equation}
  p' = 1-\phi,
\end{equation}
where $\phi$ is the normalised stellar flux during a putative extinction event. For unresolved features, the apparent radial width $W_{\rm r}$ and opacity are constrained through the finite-exposure dilution relation
\begin{equation}
  W_{\rm r}(p') = \frac{k\,\sigma_\phi\,v_{\rm rel}\,t_{\rm exp}}{p'},
\end{equation}
where $\sigma_\phi$ is the out-of-event photometric dispersion, $t_{\rm exp}$ is the effective sampling time of the series considered (Table~\ref{tab:ring}), and $k$ is the adopted detection threshold in units of $\sigma_\phi$ \citep{Santos-Sanz2022-huya,Rommel2025-huya}.
Following the standard practice in previous ring searches, we adopt $k=3$ for both the unbinned and the binned series, at all three stations.

\begin{figure}[ht!]
  \centering
  \includegraphics[width=0.85\hsize]{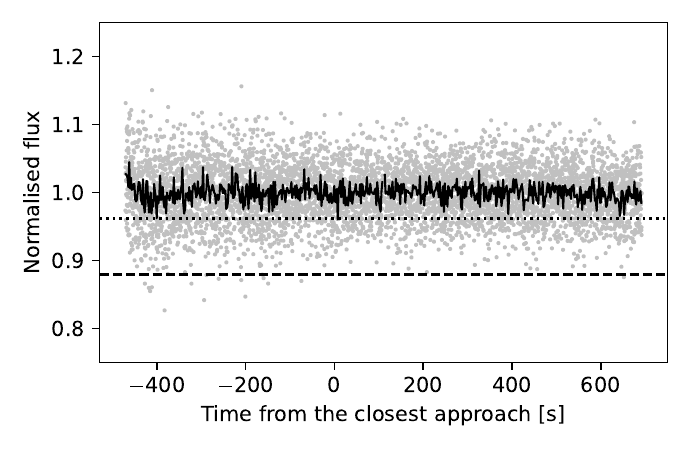}
    \caption{Normalised stellar light curve as a function of time from closest approach, obtained at the Okayama station using the 3.8~m Seimei telescope/TriCCS $r$-band camera. The grey points represent unbinned single-exposure photometry, and the black line shows fluxes binned by a factor of 10. The dashed and dotted horizontal lines correspond to the 3$\sigma$ detection levels of the unbinned and binned light curves, respectively.
     }
    \label{fig:LC_ringconst}
\end{figure}

\begin{figure}[ht!]
  \centering
  \includegraphics[width=0.9\hsize]{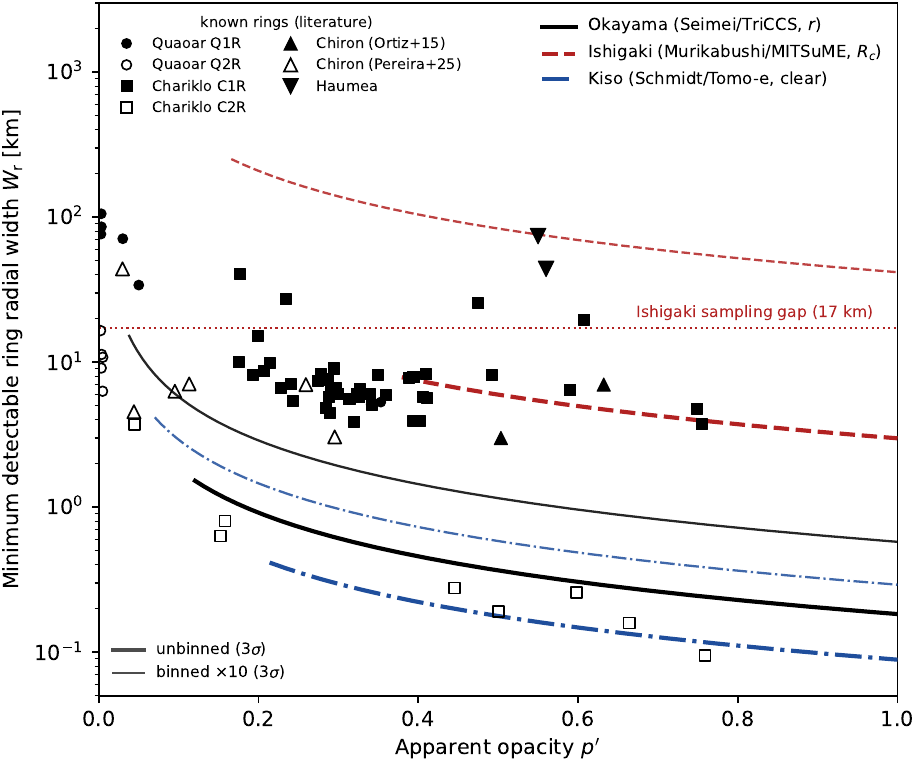}
    \caption{Observational constraints on the apparent radial width of a ring-like feature, $W_{\rm r}$, as a function of its apparent opacity, $p'$. In the present study, $p'$ is defined as $p' = 1-\phi$, where $\phi$ is the normalised stellar flux.
    Curves show the constraints obtained for the three stations: Okayama (black, solid), Ishigaki (red, dashed) and Kiso (blue, dash-dotted). For each station the thick and thin curves represent the constraints derived from the unbinned and binned light curves, respectively. The stations sample different projected distances from the body centre (about $1.1r$--$5.4r$, $3r$--$14r$ and $4r$--$20r$, respectively; see text).
    The symbols follow previous occultation analyses of small-body rings, including Chiron, Haumea, Chariklo, and Quaoar \citep{Ortiz2015-chiron,Ortiz2017-mw,Morgado2021-chariklo,Pereira2023-quaoar2rings}. Filled triangles are the two dense features detected at Chiron in 2011 \citep{Ortiz2015-chiron}, whereas open triangles are the confined rings resolved in the 2023 event \citep{Pereira2025-chiron}.
    }
    \label{fig:ring_const}
\end{figure}

Using the threshold normalised flux levels, defined as $\phi_{\rm th} = 1 - k\sigma_\phi$ (the flux level below which a single sample deviates from unity by more than $k\sigma_\phi$), $\phi_{\rm th}=0.880$ and $0.962$ for the unbinned and binned Okayama data, we obtain minimum detectable apparent opacities of $p'_{\rm min}=1-\phi_{\rm th}$, that is, $p'_{\rm min}=0.12$ (unbinned) and 0.038 (binned).
The corresponding values are 0.38 and 0.17 at Ishigaki, and 0.21 and 0.070 at Kiso. Figure~\ref{fig:LC_ringconst} shows the unbinned and binned Okayama light curves as a function of time from the closest-approach epoch inferred from the body fit, together with the corresponding detection thresholds. Converted to projected distance from Aya's best-fit shadow centre $(f_c,g_c) = (14^{+67}_{-62}~{\rm km}, 9^{+18}_{-17}~{\rm km})$, the analysed points cover approximately 430--2130~km at Ishigaki, 1260--5580~km at Okayama and 1610--8060~km at Kiso, so that the three data sets together probe about 430--8060~km ($\sim1.1r$--$20r$). No significant extinction signature is detected at any of the three stations; therefore, ring-like (or other confined circumplanetary) structures producing deeper attenuation than the derived limits are excluded over these projected distances. The resulting exclusion regions in the $(p',W_{\rm r})$ plane are summarised in Fig.~\ref{fig:ring_const}.

\begin{figure}[ht!]
  \centering
  \includegraphics[width=0.9\hsize]{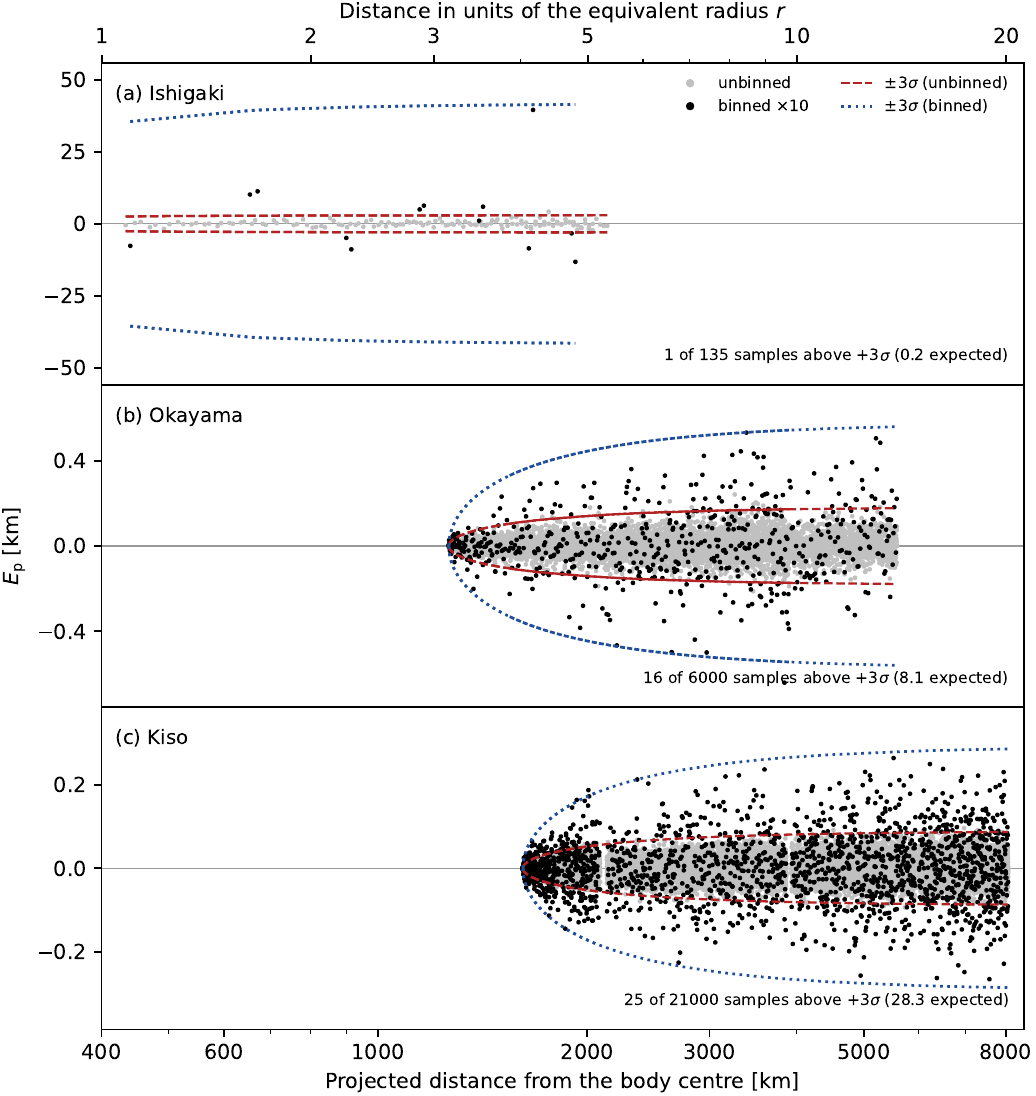}
    \caption{Equivalent width $E_{\rm p} = p'\,\Delta r$ as a function of projected distance from the fitted shadow centre for the three stations: (a) Ishigaki, (b) Okayama, and (c) Kiso, following \citet{BragaRibas2026-quaoar}. Grey and black points show the unbinned and 10-point-binned series, respectively; the red dashed and blue dotted curves are the corresponding $\pm3\sigma$ detection envelopes, which scale with the radial extent $\Delta r$ swept during one sample and therefore pinch to zero at the closest approach of each chord.
     }
    \label{fig:ring_ew}
\end{figure}

To verify that no individual excursion beyond the adopted $3\sigma$ threshold constitutes a genuine detection, we also converted each light curve from flux versus time into equivalent width versus projected distance from the fitted shadow centre, following the approach of \citet{BragaRibas2026-quaoar}.
Each sample is assigned an equivalent width
\begin{equation}
  E_{\rm p} = p'\,\Delta r,
\end{equation}
where $\Delta r = |\mathrm{d}r/\mathrm{d}t|\,t_{\rm exp}$ is the radial extent swept during a single sample.
The corresponding $1\sigma$ uncertainty is $\sigma_\phi\,\Delta r$, so the detection envelope varies with the local radial velocity of the star with respect to the shadow centre and vanishes at the closest approach of each chord.
Figure~\ref{fig:ring_ew} shows the resulting profiles for the three stations together with the $\pm3\sigma$ envelopes of the unbinned and binned series.
Of the 27135 unbinned samples, 42 exceed the $+3\sigma$ level (1, 16, and 25 at Ishigaki, Okayama, and Kiso, respectively), in overall agreement with the $\simeq37$ expected from Gaussian statistics. The binned series behave similarly (4 excursions for 3.7 expected).
The modest excess at Okayama (16 found versus 8 expected, against 5 excursions below $-3\sigma$) corresponds to a few-per-cent underestimate of $\sigma_\phi$ and is not accompanied by any correlated signal.
All excursions are confined to single samples, and no two consecutive samples exceed the threshold. The largest deviation, $4.3\sigma$, is not unexpected among $2.7\times10^{4}$ samples.
Moreover, no excursions at different stations occur at the same projected distance.
The closest pair is separated by 17~km, much larger than the radial sampling lengths in Table~\ref{tab:ring}.
We therefore conclude that all excursions beyond the $3\sigma$ level are consistent with photometric noise and that no candidate extinction event is present in any of the three data sets.

The three stations reject different classes of feature.
The Okayama and Kiso limits reject rings with apparent opacity $p'\gtrsim0.1$ and radial widths of a few kilometres or larger, comparable to the denser rings reported around Chariklo (inner dense ring C1R), Haumea, and Chiron \citep{Braga-Ribas2014-pz,Morgado2021-chariklo,Ortiz2017-mw,Ortiz2015-chiron}.
By contrast, very low-opacity rings or extremely narrow structures with $W_{\rm r}\ll1$~km cannot be ruled out, as suggested by the faint/dynamically unusual cases discussed for Quaoar and for Chariklo's outer ring C2R \citep{Pereira2023-quaoar2rings,Morgado2023-qj,Morgado2021-chariklo}, and by the evolving multi-ring system of (2060) Chiron \citep{Pereira2025-chiron}.
In their denser states ($p'\simeq0.1$--$0.3$ with $W_{\rm r}\simeq3$--$7$~km) the three confined Chiron rings lie above our exclusion curves and would have been detected in the binned Okayama and Kiso light curves, whereas in their tenuous states ($p'\lesssim0.04$), including the wide but faint configuration of Chi3R ($W_{\rm r}\simeq44$~km at $p'\simeq0.03$), they fall below the minimum detectable opacity at every station.
A Chiron-like system around Aya would therefore have been detected only if caught in a dense state.
The Ishigaki chord is the only one reaching the region immediately outside the limb, although its weaker limit constrains only nearly opaque material there. Most ring detections around small bodies are found at orbital radii of roughly $\sim3r$--$10r$ from the central body \citep{Sicardy2024-fu}.
Adopting $r\simeq401$~km for Aya, our combined 430--8060~km coverage corresponds to approximately $\sim1.1r$--$20r$, and therefore encompasses the canonical ring-search region.

  We should note that the Ishigaki observations were discontinuous with 1.0~s exposures on a 3.2~s cycle covered only 31\% of the track, leaving 17~km gaps every 25~km in which narrower features could have been missed.
  The Okayama and Kiso series were effectively continuous, except for three $\sim70$~km gaps between the Tomo-e data cubes.
  Material outside the regions sampled by the three station tracks cannot be excluded either. Continued occultation monitoring of Aya is therefore warranted, as such observations continue to reveal previously unknown material and satellites around outer solar system bodies \citep{BragaRibas2026-quaoar}.

\section{Conclusions}
\label{sec:concl}

We report the first successful stellar occultation analysis of (55565)~Aya and derive new constraints on its bulk properties and immediate environment.
Using a Bayesian fit that combines the detected chord with an ephemeris-informed prior on the shadow centre, we obtain an equivalent radius (the radius of the circle fitted to the projected shadow) of $r=401^{+28}_{-40}$~km. Together with the adopted absolute magnitude, this corresponds to a V-band geometric albedo of $p_V=0.101^{+0.023}_{-0.013}$, consistent within uncertainties with previous radiometric estimates.
The size estimate is robust against the adopted ephemeris.
 It is also marginally consistent with a fully ephemeris-independent, single-chord analysis that makes no assumption about the position of the shadow centre \citep{Morgado2026-single}, which yields $r = 341^{+59}_{-7}$~km at the 68\% credible level.
From simultaneous three-band photometry at Ishigaki, we derive 3$\sigma$ atmospheric surface-pressure upper limits of $p_0=60$, $90$, and $80$~nbar for pure CH$_4$, N$_2$, and CO compositions, respectively.
In the context of current occultation results, this provides a meaningful new atmospheric constraint in the sparsely sampled 200--500~km size regime of TNOs, and the obtained limits are lower than the 100--200~nbar level inferred for (612533)~2002~XV\textsubscript{93}.
No short-duration extinction event attributable to rings or other confined circumplanetary material is detected at any station. At Okayama, the light curve covers projected distances of 1260--5580~km ($3.2r$--$14r$) and excludes features with apparent opacity $p'\gtrsim0.1$ and radial widths of a few kilometres or larger. At Kiso, the light curve covers 1610--8060~km ($4.1r$--$20r$) and provides similar constraints on dense, kilometre-scale features. At Ishigaki, the light curve probes the innermost region, 430--2130~km ($1.1r$--$5.4r$), although its lower precision and discontinuous sampling constrain only relatively opaque features. Together, the three stations cover 430--8060~km ($1.1r$--$20r$) and no candidate extinction event is present in any data set.
However, very low-opacity or much narrower features remain possible.

These results provide the first occultation-based characterisation of Aya and demonstrate that coordinated multi-site occultation campaigns can simultaneously constrain size, albedo, atmosphere, and ring environment even when the main-body occultation is detected at a single site.

\begin{acknowledgements}
      This research has been partially supported by JSPS grants (18K13606, 21H01153).
      This research is supported by the Optical and Infrared Synergetic Telescopes for Education and Research (OISTER) programme funded by the MEXT of Japan.
      The TABASCO campaign is partially supported by the International Occultation Timing Association -- East Asia (IOTA/EA).
\end{acknowledgements}

\bibliographystyle{aa}
\bibliography{ref01}

\begin{appendix}
\nolinenumbers
\section{Comparison-star control light curves}
\label{app:control}

\begin{figure}[ht!]
  \centering
  \includegraphics[width=\hsize]{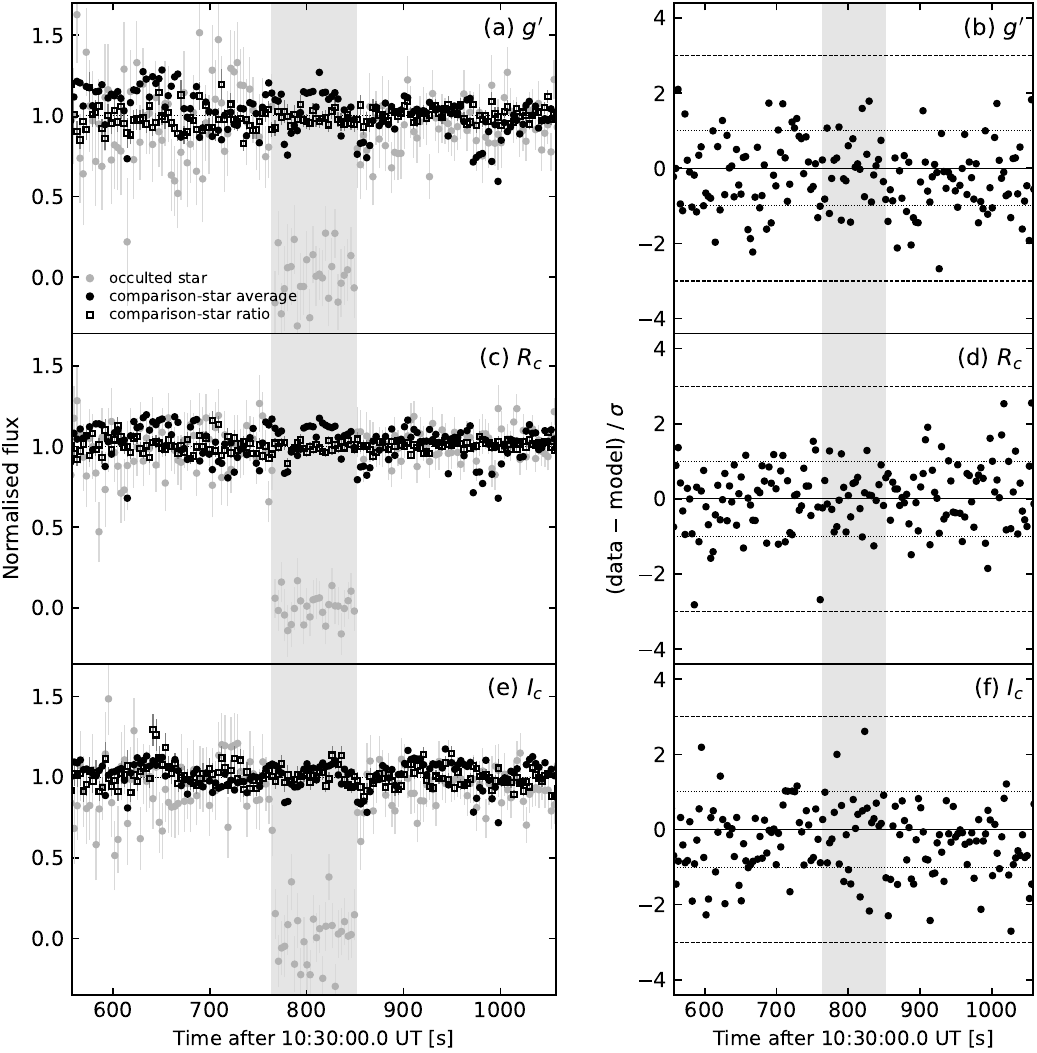}
    \caption{Comparison-star diagnostics and model residuals at the Ishigaki station, for the $g'$ (a, b), $R_c$ (c, d), and $I_c$ (e, f) bands.
    Left: normalised flux of the average of the two comparison stars (filled circles) and their mutual flux ratio (open squares), both normalised over the same baseline windows as the occulted star.
    Grey circles reproduce, for reference, the normalised fluxes of the occulted star from Fig.~\ref{fig:LC_Ishigaki}, on the same vertical scale.
    The scatter of the average reflects the atmospheric transmission, which the differential photometry divides out, while the ratio is insensitive to it. Right: residuals of the occulted star with respect to the best-fitting model, in units of the uncertainty of each sample. The dotted and dashed lines mark $\pm1\sigma$ and $\pm3\sigma$. The shaded band marks the interval between the immersion and emersion times derived in Sect.~\ref{sec:mainbody}. }
    \label{fig:control}
\end{figure}

The differential photometry at Ishigaki used two non-occulted stars observed simultaneously with the occulted star in each of the three bands (Sect.~\ref{sec:obs}).
Figure~\ref{fig:control} shows two diagnostics derived from these stars.
The left panels show the normalised flux of the average of the comparison stars, that is, the signal-to-noise-weighted reference flux by which the occulted star is divided, together with the mutual flux ratio of the two stars.
The right panels show the residuals of the occulted star with respect to the best-fitting model of Sect.~\ref{sec:mainbody}, in units of the uncertainty of each sample.
The average and the ratio carry different information.
The average tracks the atmospheric transmission and varies by 12.8, 10.7, and 8.0\% in $g'$, $R_c$, and $I_c$, which is the variation removed by the differential photometry, whereas the ratio is insensitive to transmission and is expected to be flat.
The ratio is also independent of the normalisation applied to the occulted star, since it does not involve the weighted reference flux.

Both comparison-star curves are normalised over the same baseline windows as the occulted star, and the left panels use the same vertical scale as Fig.~\ref{fig:LC_Ishigaki}.
The mutual ratio is flat throughout the run, with dispersions of 6.6, 4.2, and 7.0\%, and remains at $0.983$, $0.997$, and $1.012$ during the occultation itself, when the flux of the occulted star drops to zero.
Neither comparison-star curve shows a counterpart, at the same epoch, to any of the sub-unity excursions of the occulted star discussed in Sect.~\ref{sec:mainbody}.
Variations in atmospheric transparency or in the image quality can therefore be excluded as their origin.

In the right panels, over all 164 samples of each band, including those acquired during the occultation, no residual exceeds $3\sigma$.
The most deviant samples reach $-2.7$, $-2.8$, and $-2.7\sigma$ in $g'$, $R_c$, and $I_c$. The excursions are thus accounted for by the photon and detector noise of the much fainter occulted star, and no residual structure suggests an unmodelled extinction event.

\end{appendix}

\end{document}